\documentclass[aps, reprint, prl, superscriptaddress]{revtex4-2}
\usepackage{newtxtext, newtxmath}
\usepackage{dcolumn}
\usepackage{braket}
\usepackage{graphicx}
\usepackage{color}
\usepackage{orcidlink}
\definecolor{bblue}{rgb}{0, 0.0, 0.8}
\definecolor{rred}{rgb}{0.7, 0.0, 0.0}

\usepackage{hyperref}
\hypersetup{
    colorlinks=true,        
    linkcolor=bblue,          
    citecolor=bblue,        
    filecolor=bblue,      
    urlcolor=bblue           
}

\begin{document}
\title{Vector pulse magnet}
\author{Kosuke Noda}
\author{Kenta Seki}
\author{Dilip Bhoi\orcidlink{0000-0001-7333-7286}}
\author{Kazuyuki Matsubayashi\orcidlink{0000-0002-8814-1910}}
\affiliation{Department of Engineering Science, University of Electro-Communications, Chofu, Tokyo 182-8585, Japan}
\author{Kazuto Akiba\orcidlink{0000-0003-1408-4922}}
\affiliation{Faculty of Science and Engineering, Iwate University, Morioka, Iwate
020-8551, Japan}
\author{Akihiko Ikeda\orcidlink{0000-0001-7642-0042}}
\email[]{a-ikeda@uec.ac.jp}
\affiliation{Department of Engineering Science, University of Electro-Communications, Chofu, Tokyo 182-8585, Japan}

\date{\today}

\begin{abstract}
The underlying symmetry of the crystal, electronic structure, and magnetic structure manifests itself in the anisotropy of materials' properties, which is a central topic of the present condensed matter research.
However, it demands such a considerable effort to fill the explorable space that only a small part has been conquered.
We report a vector pulse magnet (VPM) as an alternative experimental technique to control the direction of applied magnetic fields, which may complement the conventional methods with its characteristic features.
The VPM combines a conventional {\it pulse} magnet and a {\it vector} magnet.
The VPM can create vector pulsed magnetic fields and swiftly rotating pulsed magnetic fields.
As a demonstration, the three-dimensional magnetoresistance measurement of a highly oriented pyrolytic graphite is carried out using the AC four-probe method at 4.5 K and 6 T.
The two-dimensional electronic structure of graphite is visualized in the three-dimensional magnetoresistance data.
One can uncover the rotational and time-reversal symmetry of materials using a VPM and a variety of measurement techniques.
\end{abstract}

\maketitle

\begin{figure}
\begin{center}
\includegraphics[width =0.85\columnwidth]{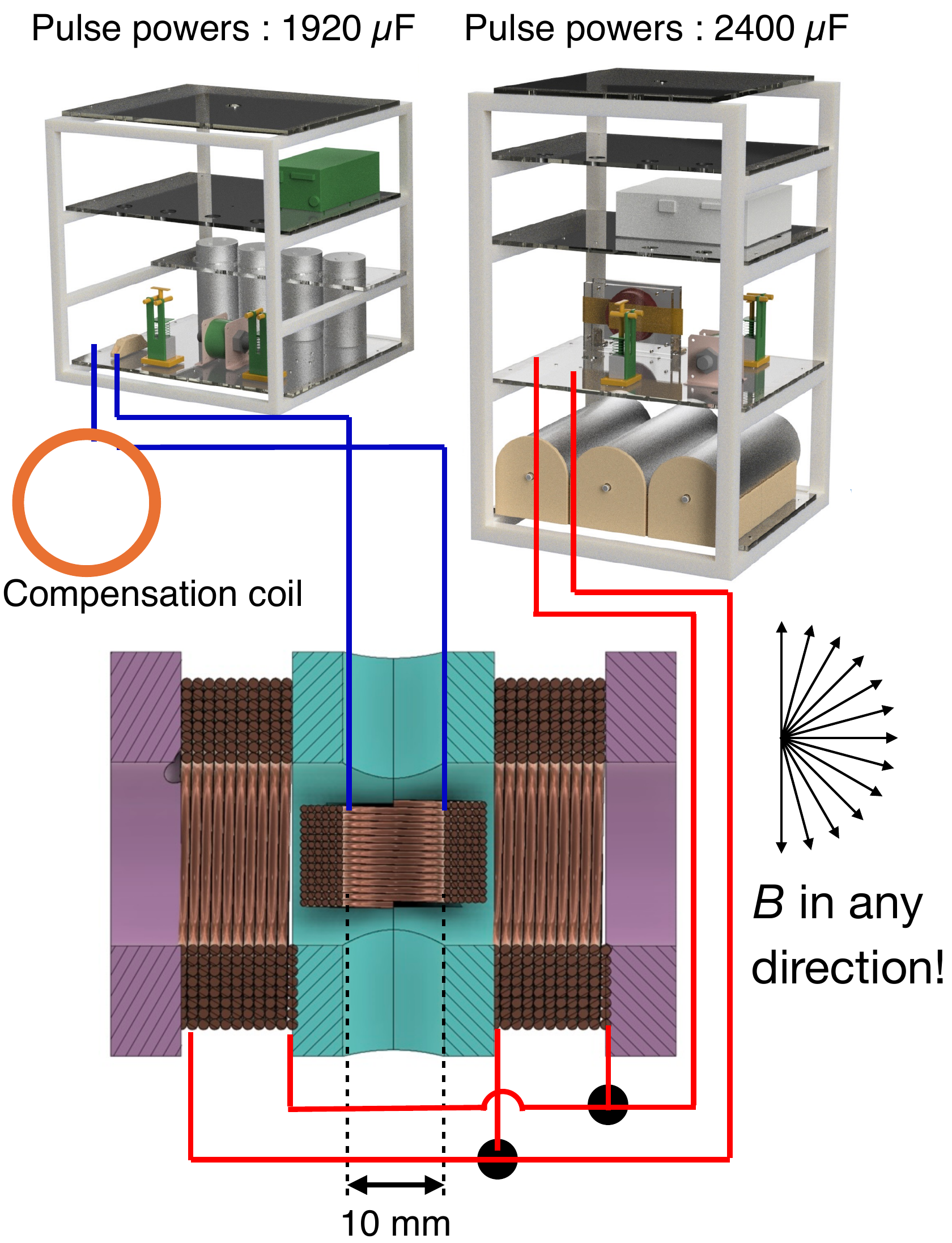}
\caption{
A schematic drawing of the VPM that can create a magnetic field in an arbitrary direction.
The VPM is made of an inner coil and an outer coil.
The inner and the outer coils are a mono-coil and a split-pair coil, respectively.
One power supply is connected to the inner coil.
The other power supply is connected to the outer split pair coil.
The two electric circuits are independent, so one can trigger them with arbitrary delayed timing.
\label{vpm}}
\end{center}
\end{figure}

\begin{figure}
\begin{center}
\includegraphics[width =0.8\columnwidth]{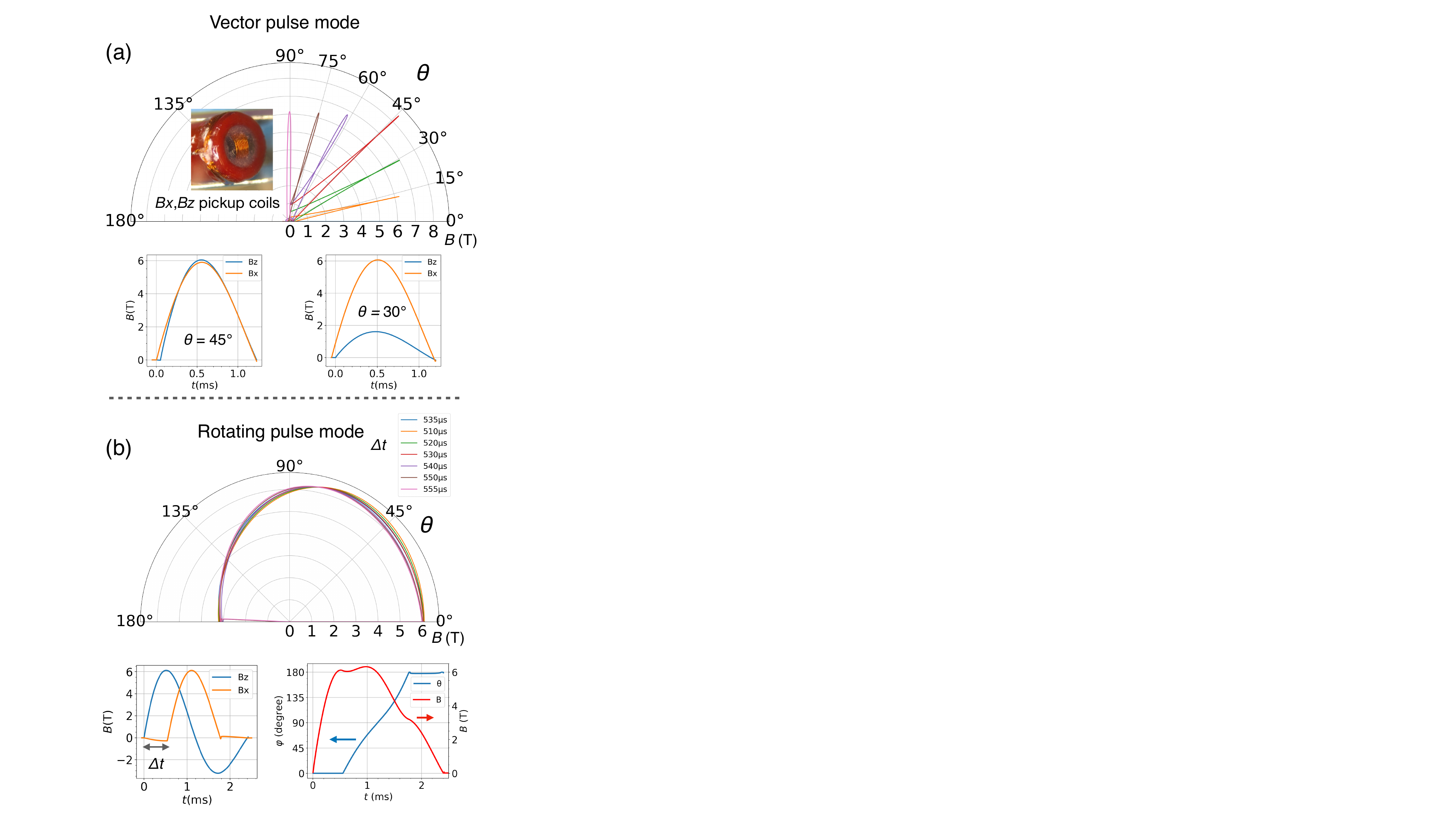}
\caption{
(a) The trajectory of the generated vector pulsed magnetic field using a VPM plotted on the polar coordinates.
The two insets show the magnetic field profiles generated by the inner coil $B_{z}$, and the outer coil $B_{x}$.
Notice that two pulses are generated at the same time.
The ratio of the magnetic field strength $B''/B'$ determines the direction $\theta$ of the generated magnetic field.
The photo shows two orthogonal pickup coils for simultaneous measurements of $B_{x}$ and $B_{z}$.
(b) The trajectory of the generated rotating pulsed magnetic field using a VPM plotted on the polar coordinates.
The inset at bottom left shows the magnetic field profiles generated by the inner coil $B_{z}$, and the outer coil $B_{x}$.
Notice that the pulses are shifted for $\pi/4$ in phase from each other.
The inset at bottom right shows the profiles of $|B|$ and $\theta$ generated by the VPM.
\label{vp}}
\end{center}
\end{figure}

Anisotropy in the properties of solids reflects the underlying symmetry of their electronic and magnetic structures, including features such as topological electronic states and exotic orders found in correlated electron systems and quantum magnets \cite{YonezawaNP2017, NiiPRL2025, NomuraPRL2019, NomuraPRL2023, FukushimaNC2024, AsabaNP2024, FangPRL2025, ImamuraSA2025}.
Crystalline materials exhibit anisotropy in various physical properties, such as elastic modulus, dielectric constant, and magnetization, as well as in transport phenomena like electrical resistivity, the Hall effect, and the thermal Hall effect.
According to Neumann’s principle, the macroscopic material tensors must be invariant under the symmetry operations of the crystal’s point group. This symmetry can be further modified or enriched by magnetic ordering and spin-dependent electronic dispersions.
Such off-diagonal responses in materials are particularly intriguing in noncollinear antiferromagnets \cite{NakatsujiNature2015, KimataNature2019, QinNature2023} and altermagnets \cite{FengNatElectron2022, GonzalezPRL2023}, due to their potential applications with negligibly small stray magnetic fields.

A magnetic field is a unique stimulus for visualizing the inherent symmetry of the material.
The magnetic field is an axial vector field that breaks the material's time-reversal symmetry by directly affecting the electrons' spin and orbital magnetic moment.
Here, we devised an alternative technique, a vector pulse magnet (VPM),  to control the direction of the magnetic field.
The idea of VPM is to make a {\it pulse} version of the conventional {\it vector} magnet (VM).
We devised a VPM composed of two pulse magnets that generate a magnetic field perpendicular to each other, as shown in Fig. \ref{vpm}.
Two independent pulse powers drive the outer and inner coils.
The outer coil is a split pair coil with an 18 mm bore, and a mono-coil with a 10 mm bore is installed as an inner coil.
VPM generates a pulsed magnetic field in an arbitrary direction.
With VPM, one can investigate the material's rotational and time-reversal symmetry.
VPM has different natures compared to the conventional techniques, such as a VM and a two-axis rotation stage in a DC magnet.
VPM operates in either of the two modes as follows.

\paragraph*{Vector pulse mode:}
Pulsed magnetic fields are generated at an arbitrary angle $\theta$ with a single pulse of a few milliseconds.
The vector pulse is qualitatively described as
\begin{equation}
\mathbf{B}_{{\rm vect}}(t) =
\begin{pmatrix}
 B_{x} \\
 B_{y} \\
 B_{z} \\
\end{pmatrix}
 = 
 \begin{pmatrix}
B'\sin 2\pi t/\tau \\
0 \\
B''\sin 2\pi t/\tau \\
\end{pmatrix},
\end{equation}
$\theta$ is controlled by the relative intensity of each pulsed magnetic field $\theta = \tan^{-1}{(B''/B')}$.
The obtained trajectories of the vector magnetic field are shown in Fig. \ref{vp}(a).
The vector magnetic field is generated in a pulse with a variety of $\theta$ from 0$^{\circ}$ to 90$^{\circ}$.
The strength of the magnetic field $|B| = \sqrt{B'^{2} + B''^{2}}$ is variable by changing the charging voltage, where the maximum strength is $\sqrt{2}$ times larger at $\theta = 45^{\circ}$ than those at $\theta = 0^{\circ}$ and $90^{\circ}$.
The trajectories of the vector magnetic field show a slight open loop in Fig. \ref{vp}(a), which is not significant at $B_{\rm max} / 2 <B<B_{\rm max}$, but is significant at $B < B_{\rm max} / 2$. 
Hence, to obtain a directed magnetic field at low magnetic fields, we need additional pulses with smaller $B_{\rm max}$.

\paragraph*{Rotating pulse mode:}
Pulsed magnetic fields rotating from $\theta = 0^{\circ}$ to a greater $\theta$ in a single pulse of a few milliseconds are generated.
The rotation happens as
\begin{equation}
\mathbf{B}_{{\rm rot}}(t) =
\begin{pmatrix}
 B_{x} \\
 B_{y} \\
 B_{z} \\
\end{pmatrix}
 = 
 \begin{pmatrix}
B_{0}\sin 2\pi t/\tau \\
0 \\
B_{0}\cos 2\pi t/\tau \\
\end{pmatrix},
\end{equation}
which is realized by delaying the start of the pulse of $B_{x}$ from the preceding pulse $B_{z}$ for half the pulse duration $\tau/4$, where $B_{0}$ is the constant magnetic field we want to rotate.
As shown in the two plots at the bottom of Fig. \ref{vp}(b), $B_{x}$ and $B_{z}$ are generated with the same pulse duration of $\sim1$ ms and the delay of $\sim0.5$ ms.
The resultant trajectories of the rotating magnetic field are shown at the top of Fig. \ref{vp}(b) with a variation of the delay, where the magnetic field is rotated from $\theta=0^{\circ}$ to 180$^{\circ}$.
Note that the strength of the magnetic field is kept at $\sim6$ T from $\theta=0^{\circ}$ to 90$^{\circ}$, followed by a drastic decrease to 3 T from $\theta=90^{\circ}$ to 180$^{\circ}$.
The time duration in the rotating pulse from $\theta=0^{\circ}$ to 90$^{\circ}$ is  $\sim1$ ms.
The strength of the magnetic field is variable by coherently changing the charging voltages of each pulse power.

Both vector pulse magnetic field and rotating pulse magnetic field are reproducible and stable.
The heating of the coil is insignificant, so one can repeat the magnetic field pulse every 2 min.

\begin{figure}
\begin{center}
\includegraphics[width =0.9\columnwidth]{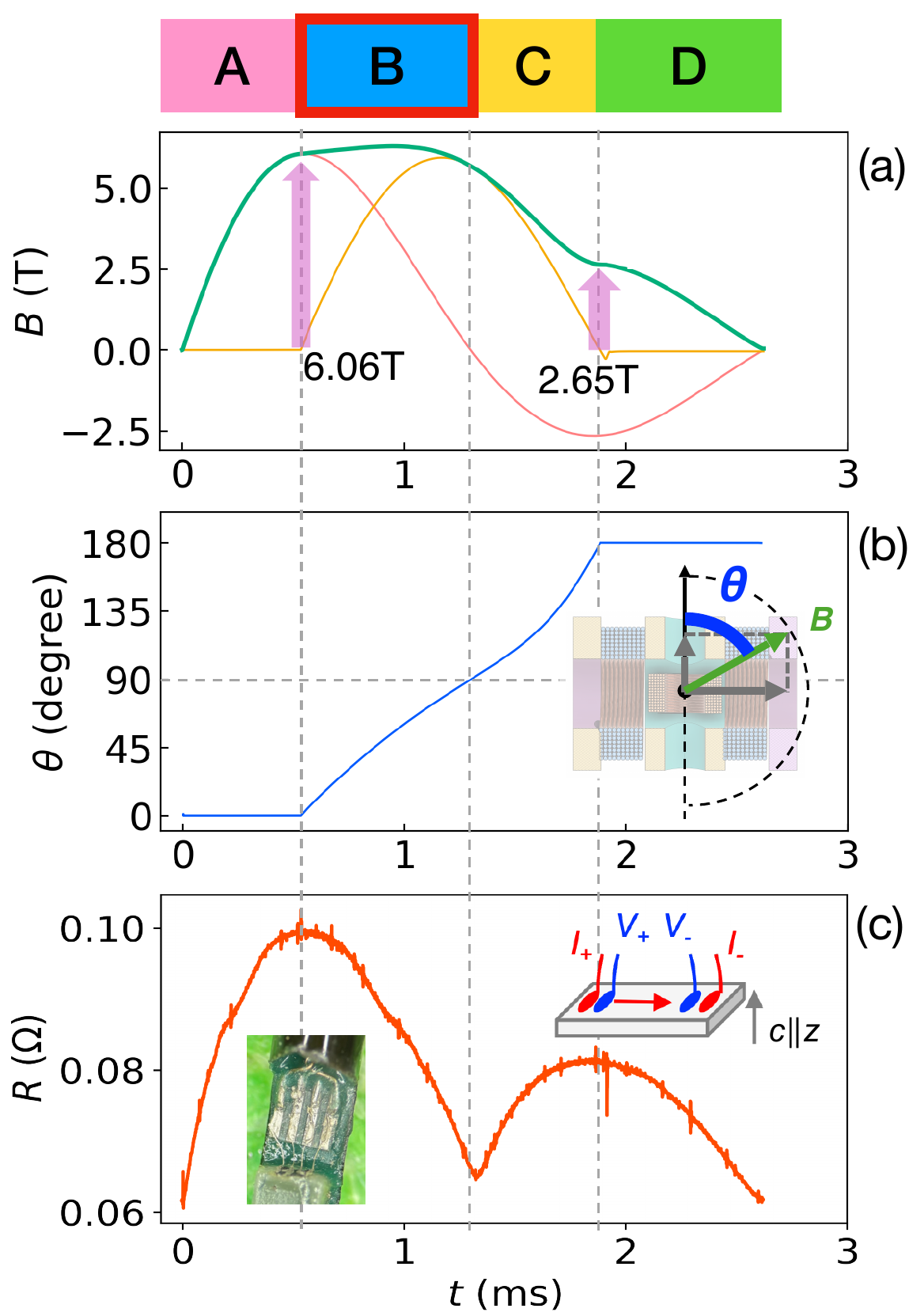}
\caption{
(a) The strength of the rotating magnetic field $|B| = \sqrt{B_z^{2} + B_{xy}^{2}}$ as a function of time, shown with the magnetic field profile of the inner coil $B_z$, and outer coil $B_{xy}$.
(b) The angle of the rotating magnetic field $\theta = \tan^{-1}(B_{z}/B_{xy})$ as a function of time.
(c) The $R_{xx}$ data of HOPG shown as a function of time measured using the AC four-probe method.
\label{rot}}
\end{center}
\end{figure}

To demonstrate the rotating magnetic field generated using the VPM, we measured the magnetoresistance of a highly oriented pyrolytic graphite (HOPG) at 4.5 K.
HOPG is a suitable test sample due to its large magnetoresistance and strong two-dimensional electronic structure \cite{BarzolaPRM2019, IkedaJAP2024}.
In Figs. \ref{rot}(a) and \ref{rot}(b) , we generate a rotating magnetic field keeping $|B|=6$ T at $\theta$ between $0^{\circ}$ and 90$^{\circ}$.
On the other hand, we obtained $|B|<6$ T between $90^{\circ}$ and 180$^{\circ}$
The magnetoresistance $R_{xx}(\theta, B)$  measurement is carried out under the rotating pulsed magnetic field as shown in Figs. \ref{rot}(c).
The magnetic field grows in the time region $A$ and is applied only in the $z$ direction, $\theta = 0$.
Here, the magnetic field is parallel to the $c$ axis.
The resistance data $R_{xx}$ show a positive magnetoresistance, which increases with increasing magnetic field, agreeing with previous reports on HOPG \cite{BarzolaPRM2019, IkedaJAP2024}.
In the time region $B$, the magnetic field is rotating from $\theta = 0^{\circ}$ to $90^{\circ}$ keeping the field strength at $|B|\simeq6$ T.
In the time region $C$, the magnetic field keeps rotating from $\theta = 90^{\circ}$ to $180^{\circ}$.
Here, the magnetic field strength decreases from $|B| = 6$ T to 2.65 T.
Notice that $R_{xx}$ keeps decreasing in the time region $B$ and starts increasing in the time region $C$.
This indicates that the magnetoresistance is sensitive to the component of the magnetic field applied in parallel to the $c$ axis of HOPG.
In the time region $D$, the magnetic field is applied only in the $z$ direction, $\theta = 180^{\circ}$, which decreases from 2.65 T to 0 T.
The present VPM generates the rotating magnetic fields with the variation of $|B|$ in the time region $B$ of  $\pm 3\%$, which is already useful and would be improved.
On the other hand, we find it more challenging to keep $|B|$ in the time region $C$.

\begin{figure}
\begin{center}
\includegraphics[width =\columnwidth]{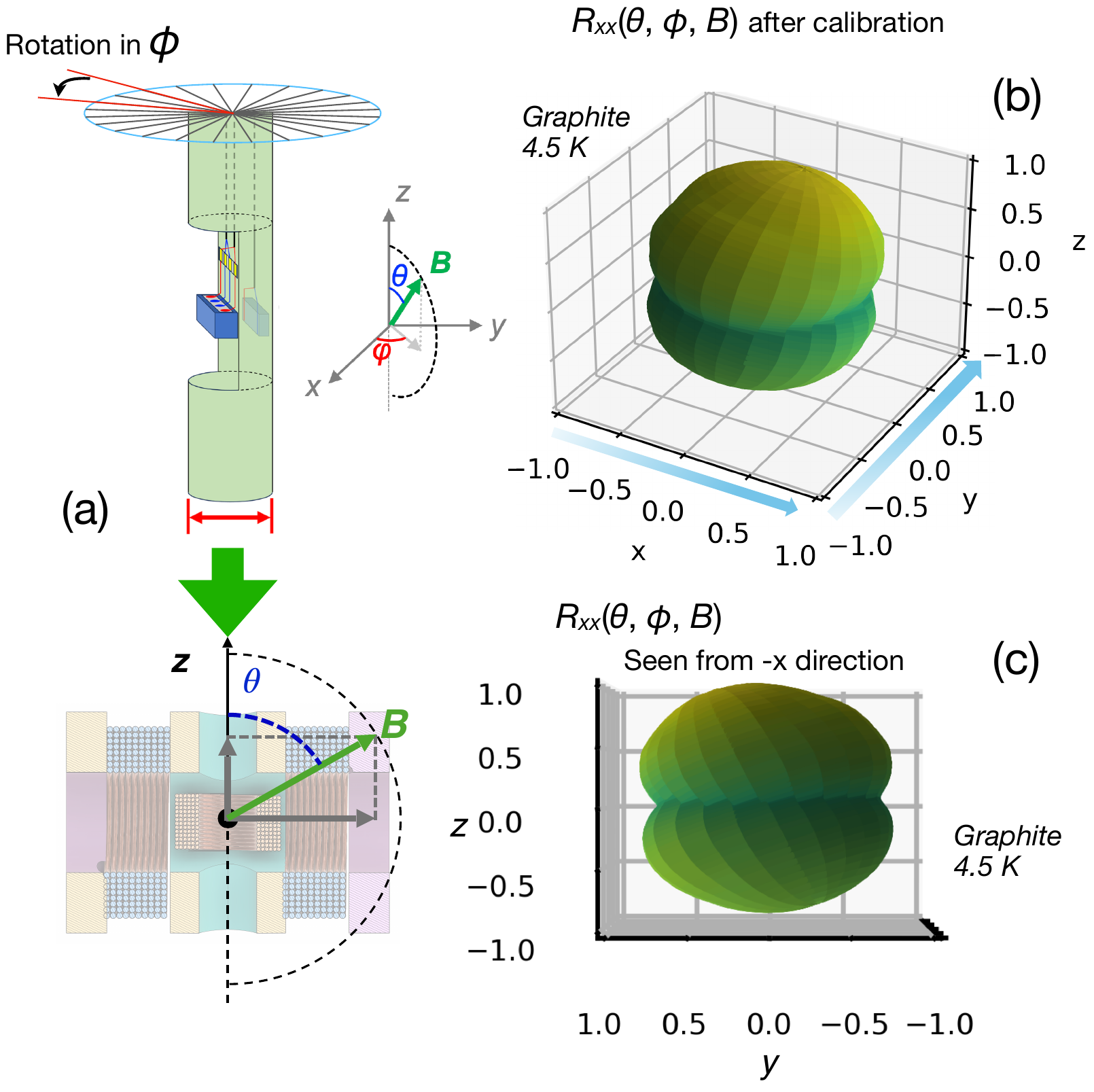}
\caption{
(a) A schematic to show how to change $\phi$ by rotating the sample probe.
(b) $R_{xx}(\theta, \phi, B)$ mapped in 3D space,
where $B \simeq 6$ T for $0^{\circ}<\theta<90^{\circ}$ and $B< 6$ T for $90^{\circ}<\theta<180^{\circ}$.
(c) Side view of the same data seen from $-x$ direction.
\label{result1}}
\end{center}
\end{figure}

We further conducted multiple measurements to visualize the magnetoresistive feature with magnetic field direction in the entire $4\pi$ solid angle, where we rotated the sample probe by $\Delta\phi = 15^{\circ}$ from shot to shot.
We collected 24 data in total, which took us $\sim1$ hour.
The collection of the data, $R_{xx}(\theta, \phi, B)$, is mapped to the orthogonal 3D space using the following relation as shown in Fig. \ref{result1}.
\begin{equation}
\begin{pmatrix}
x \\
y \\
z \\
\end{pmatrix}
=
\begin{pmatrix}
R_{xx}(\theta, \phi, B)\sin\theta\cos\phi \\
R_{xx}(\theta, \phi, B)\sin\theta\sin\phi \\
R_{xx}(\theta, \phi, B)\cos\theta \\
\end{pmatrix}.
\end{equation}
Note that $B \simeq 6$ T for $0^{\circ}<\theta<90^{\circ}$ and $B< 6$ T for $90^{\circ}<\theta<180^{\circ}$.
The original data contains a misalignment between the sample normal ($c$ axis) and the external magnetic field of $B_{z}$.
We used the following rotation matrix to calibrate the misalignment in the data.
\begin{equation}
\begin{pmatrix}
x' \\
y' \\
z' \\
\end{pmatrix}
=
\begin{pmatrix}
\cos \beta & 0 &  -\sin \beta \\
0 & 1 & 0 \\
\sin \beta & 0 & \cos \beta \\
\end{pmatrix}
\begin{pmatrix}
\cos \alpha & \sin \alpha & 0  \\
-\sin \alpha & \cos \alpha & 0 \\
0 & 0 & 1 \\
\end{pmatrix}
\begin{pmatrix}
x \\
y \\
z \\
\end{pmatrix},
\end{equation}
with $\alpha$ and $\beta$ being the Euler's angles for calibration.
The obtained data is plotted as a curved surface in Fig. \ref{result1}(b).
The side view of the same data is plotted in Fig. \ref{result1}(c).

\begin{figure}
\begin{center}
\includegraphics[width =\columnwidth]{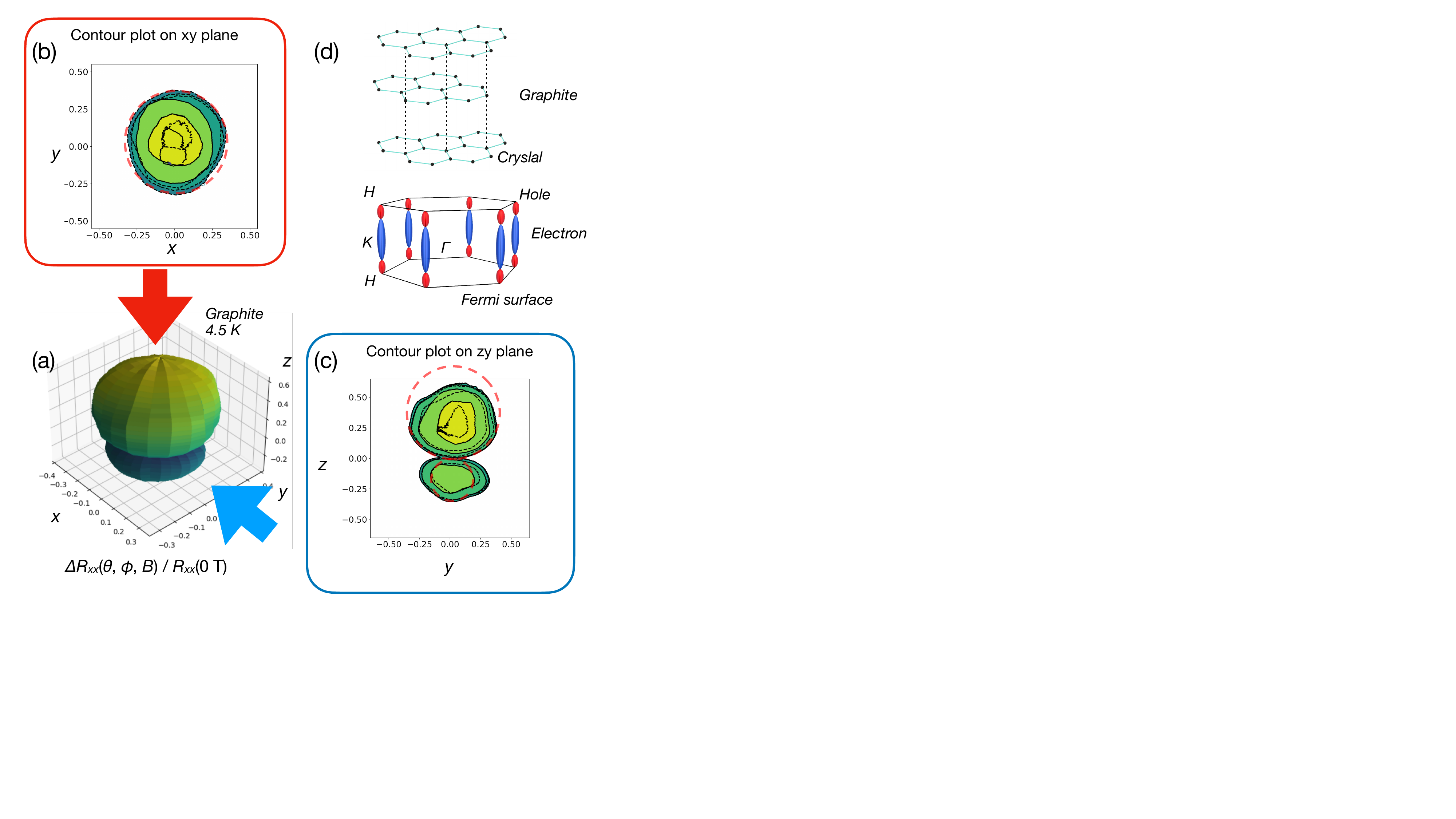}
\caption{
(a) The data of $\Delta R_{xx}(\theta, \phi, B)/R_{xx}(0 {\rm\ T})$ mapped in the 3D space.
Here, note that $B \simeq 6$ T for $0^{\circ}<\theta<90^{\circ}$ and $B< 6$ T for $90^{\circ}<\theta<180^{\circ}$.
(b) The contour plot of the data in the $xy$ plane and (c) $zy$ plane.
(d) The schematic drawings of the crystal structure and Fermi surface of graphite. 
\label{result2}}
\end{center}
\end{figure}

In Fig. \ref{result2}(a), we show the 3D data of the magnetoresistance $\Delta R_{xx}(\theta, \phi, B)/R_{xx}(0 {\rm\ T})$, where $\Delta R_{xx}(\theta, \phi, B) = R_{xx}(\theta, \phi, B) - R_{xx}(0 {\rm\ T})$.
Here again, note that $B \simeq 6$ T for $0^{\circ}<\theta<90^{\circ}$ and $B< 6$ T for $90^{\circ}<\theta<180^{\circ}$.
The contour plots of the magnetoresistance on the $xy$ plane and $yz$ plane are shown in Figs. \ref{result2}(b) and \ref{result2}(c).
The crystal structure and the schematic drawing of the Fermi surface of graphite is shown in Fig. \ref{result2}(d).
$\Delta R_{xx}(\theta, \phi, B)/R_{xx}(0 {\rm\ T})$ shows a shape of two balls connected at the origin.
Each ball is slightly compressed in the $z$ direction.
The top ball at $z>0$ is smaller than the bottom at $z<0$.
The contour plot on the $xy$ plane shows that the ball is almost circular when viewed from the top side.

We give a qualitative discussion on the result.
The magnetoresistance of graphite may be expressed as $\Delta R_{xx}(\theta, \phi, B) / R_{xx}(0 {\rm\ T}) \propto 1/m_c^e m_c^h$, where $m_{c}^{e}$ and $m_{c}^{h}$ are the cyclotron masses of electron and hole pockets, respectively, and the two-carrier model is employed because graphite is a compensated semi-metal  \cite{AkibaJPSJ2015, AkibaPRB2017}.
The Fermi surface of graphite shown in the inset of Fig. \ref{result2} indicates that the anisotropy of the effective mass tensor $m_{zz} >> m_{0} >> m_{xx}\sim m_{yy}$ for both electron hole and pockets.
With $B\parallel c$ and $B\perp c$, the cyclotron masses are expressed as $\sqrt{m_{xx}m_{yy}}$ and $\sqrt{m_{xx(yy)}m_{zz}}$, respectively, where the former is light and the latter is heavy because the latter includes $m_{zz}$.
They make the magnetoresistance considerable with $B\parallel c$ and negligible with $B\perp c$.
Hence, the expected shape of the magnetoresistance would be $\cos\theta$, which is indicated by the red dashed lines on the contour plots.
It shows a good agreement on the contour plot in Fig. \ref{result2}(b).
On the other hand, we see the deviation from the $\cos\theta$ circle in Fig. \ref{result2}(c).
The reason for the compression of the circle is the deviation from the linear magnetoresistance above 3 T as reported in Ref. \cite{IkedaJAP2024}, which should result in the reduced magnetoresistance at 6 T compared to the $\cos\theta$ circle.
The reason for the smaller circle in $z<0$ is extrinsic.
The magnetic field strength is not sustained in the time region $C$ in Fig. \ref{rot}(c).
A quick solution to acquire data for $90^{\circ}<\theta<180^{\circ}$ at 6 T is to reverse the polarity of the power cable by reconnecting it accordingly.

Here, we compare our VPM with conventional techniques, namely, VM and a two-axis mechanical rotator method, from several viewpoints.
Although high-precision measurements are currently best performed using VM and mechanical rotator methods, we aim to highlight measurements that are uniquely accessible with VPM.
(i) A rigid stick-type probe is usable.
From the viewpoints of measurements, not only the magnetoresistance measurements in the present study, but also the measurements of Hall effect, magneto-optical effect, and magnetostriction using an optical fiber \cite{IkedaRSI2017, IkedaRSI2018, IkedaPRB2019} are applicable.
Note that the measurements with optical fibers are suitable for VPM and VM, which does not apply to the mechanical rotator method.
(ii) Pulse-favored measurements are applicable.
The induction magnetization measurement ($dM/dt$) and pyroelectric current measurement ($dC/dt$) are readily applicable in VPM owing to the pulse generation.
With a VPM in the {\it vector} pulse mode, one obtains $(\partial M / \partial B)_{\theta, \phi}$ which gives a magnetization curve as a function of magnetic field in arbitrary direction $M(B)_{\theta, \phi}$.
With a VPM in the {\it rotating} pulse mode, one obtains $(\partial M / \partial \theta)_{B, \phi}$ which gives a magnetization curve as a function of $\theta$ at a constant magnetic field $M(\theta)_{B, \phi}$.
These measurements are only available in VPM.
(iii). Possible improvements in $B_{\rm max}$.
In the present study, we prototyped a VPM with a typical magnetic field strength of 6 T, which already exceeds that of VM by a factor of two.
We estimate that an improved VPM should have a larger magnetic field, approaching 10 T, based on our experience in generating fields exceeding 40 T using a mono-coil and the same pulse power \cite{IkedaJAP2024}.
However, as for $B_{\rm max}$, the mechanical rotor method is applicable up to 40 T DC magnet in specialized facilities.
In comparison, VPM is a laboratory-scale setup that is much more accessible.

In summary, we devised a VPM that generates either a vector-pulsed magnetic field or a rotating-pulse magnetic field of up to 6 T by adjusting the pulse's delay.
The VPM consists of two pulse magnets with distinct pulse powers whose magnetic fields are oriented orthogonally to each other.
As a demonstration, the 2D nature of HOPG's magnetoresistance is visualized in 3D space at 6 T, which reveals that the electronic structure of the Fermi surface is two-dimensional.
VPM can be an alternative approach to reveal the manifestation of microscopic symmetry in materials' macroscopic data, by further exploiting the characteristics of the VPM such as the study of non-equilibrium study of materials using the fast sweeping rotation of the magnetic field and also the possible achievement of 10 T which is not reachable using a conventional VM.

\begin{acknowledgments}
This work is supported by the JST FOREST program No. JPMJFR222W, JSPS Grant-in-Aid for Scientific Research on Innovative Areas (A) \textquotedblleft 1000 T SCIENCE\textquotedblright  \ , 23H04859, 23H04861, 23H04862, and Grant-in-Aid for Scientific Research (B) 23H01121, and MEXT LEADER program No. JPMXS0320210021.
\end{acknowledgments}
\bibliography{vpm}

\begin{thebibliography}{20}%
\makeatletter
\providecommand \@ifxundefined [1]{%
 \@ifx{#1\undefined}
}%
\providecommand \@ifnum [1]{%
 \ifnum #1\expandafter \@firstoftwo
 \else \expandafter \@secondoftwo
 \fi
}%
\providecommand \@ifx [1]{%
 \ifx #1\expandafter \@firstoftwo
 \else \expandafter \@secondoftwo
 \fi
}%
\providecommand \natexlab [1]{#1}%
\providecommand \enquote  [1]{``#1''}%
\providecommand \bibnamefont  [1]{#1}%
\providecommand \bibfnamefont [1]{#1}%
\providecommand \citenamefont [1]{#1}%
\providecommand \href@noop [0]{\@secondoftwo}%
\providecommand \href [0]{\begingroup \@sanitize@url \@href}%
\providecommand \@href[1]{\@@startlink{#1}\@@href}%
\providecommand \@@href[1]{\endgroup#1\@@endlink}%
\providecommand \@sanitize@url [0]{\catcode `\\12\catcode `\$12\catcode
  `\&12\catcode `\#12\catcode `\^12\catcode `\_12\catcode `\%12\relax}%
\providecommand \@@startlink[1]{}%
\providecommand \@@endlink[0]{}%
\providecommand \url  [0]{\begingroup\@sanitize@url \@url }%
\providecommand \@url [1]{\endgroup\@href {#1}{\urlprefix }}%
\providecommand \urlprefix  [0]{URL }%
\providecommand \Eprint [0]{\href }%
\providecommand \doibase [0]{https://doi.org/}%
\providecommand \selectlanguage [0]{\@gobble}%
\providecommand \bibinfo  [0]{\@secondoftwo}%
\providecommand \bibfield  [0]{\@secondoftwo}%
\providecommand \translation [1]{[#1]}%
\providecommand \BibitemOpen [0]{}%
\providecommand \bibitemStop [0]{}%
\providecommand \bibitemNoStop [0]{.\EOS\space}%
\providecommand \EOS [0]{\spacefactor3000\relax}%
\providecommand \BibitemShut  [1]{\csname bibitem#1\endcsname}%
\let\auto@bib@innerbib\@empty
\bibitem [{\citenamefont {Yonezawa}\ \emph {et~al.}(2017)\citenamefont
  {Yonezawa}, \citenamefont {Tajiri}, \citenamefont {Nakata}, \citenamefont
  {Nagai}, \citenamefont {Wang}, \citenamefont {Segawa}, \citenamefont {Ando},\
  and\ \citenamefont {Maeno}}]{YonezawaNP2017}%
  \BibitemOpen
  \bibfield  {author} {\bibinfo {author} {\bibfnamefont {S.}~\bibnamefont
  {Yonezawa}}, \bibinfo {author} {\bibfnamefont {K.}~\bibnamefont {Tajiri}},
  \bibinfo {author} {\bibfnamefont {S.}~\bibnamefont {Nakata}}, \bibinfo
  {author} {\bibfnamefont {Y.}~\bibnamefont {Nagai}}, \bibinfo {author}
  {\bibfnamefont {Z.}~\bibnamefont {Wang}}, \bibinfo {author} {\bibfnamefont
  {K.}~\bibnamefont {Segawa}}, \bibinfo {author} {\bibfnamefont
  {Y.}~\bibnamefont {Ando}},\ and\ \bibinfo {author} {\bibfnamefont
  {Y.}~\bibnamefont {Maeno}},\ }\bibfield  {title} {\bibinfo {title} {\rm{
  Thermodynamic evidence for nematic superconductivity in
  Cu$_{x}$Bi$_{2}$Se$_{3}$}},\ }\href {https://doi.org/10.1038/nphys3907}
  {\bibfield  {journal} {\bibinfo  {journal} {Nat. Phys.}\ }\textbf {\bibinfo
  {volume} {13}},\ \bibinfo {pages} {123} (\bibinfo {year} {2017})}\BibitemShut
  {NoStop}%
\bibitem [{\citenamefont {Nii}\ \emph {et~al.}(2025)\citenamefont {Nii},
  \citenamefont {Yamamoto}, \citenamefont {Kanno}, \citenamefont {Maekawa},\
  and\ \citenamefont {Onose}}]{NiiPRL2025}%
  \BibitemOpen
  \bibfield  {author} {\bibinfo {author} {\bibfnamefont {Y.}~\bibnamefont
  {Nii}}, \bibinfo {author} {\bibfnamefont {K.}~\bibnamefont {Yamamoto}},
  \bibinfo {author} {\bibfnamefont {M.}~\bibnamefont {Kanno}}, \bibinfo
  {author} {\bibfnamefont {S.}~\bibnamefont {Maekawa}},\ and\ \bibinfo {author}
  {\bibfnamefont {Y.}~\bibnamefont {Onose}},\ }\bibfield  {title} {\bibinfo
  {title} {\rm{Observation of Nonreciprocal Diffraction of Surface Acoustic
  Wave}},\ }\href {https://doi.org/10.1103/PhysRevLett.134.027001} {\bibfield
  {journal} {\bibinfo  {journal} {Phys. Rev. Lett.}\ }\textbf {\bibinfo
  {volume} {134}},\ \bibinfo {pages} {027001} (\bibinfo {year}
  {2025})}\BibitemShut {NoStop}%
\bibitem [{\citenamefont {Nomura}\ \emph {et~al.}(2019)\citenamefont {Nomura},
  \citenamefont {Zhang}, \citenamefont {Zherlitsyn}, \citenamefont {Wosnitza},
  \citenamefont {Tokura}, \citenamefont {Nagaosa},\ and\ \citenamefont
  {Seki}}]{NomuraPRL2019}%
  \BibitemOpen
  \bibfield  {author} {\bibinfo {author} {\bibfnamefont {T.}~\bibnamefont
  {Nomura}}, \bibinfo {author} {\bibfnamefont {X.-X.}\ \bibnamefont {Zhang}},
  \bibinfo {author} {\bibfnamefont {S.}~\bibnamefont {Zherlitsyn}}, \bibinfo
  {author} {\bibfnamefont {J.}~\bibnamefont {Wosnitza}}, \bibinfo {author}
  {\bibfnamefont {Y.}~\bibnamefont {Tokura}}, \bibinfo {author} {\bibfnamefont
  {N.}~\bibnamefont {Nagaosa}},\ and\ \bibinfo {author} {\bibfnamefont
  {S.}~\bibnamefont {Seki}},\ }\bibfield  {title} {\bibinfo {title} {Phonon
  magnetochiral effect},\ }\href
  {https://doi.org/10.1103/PhysRevLett.122.145901} {\bibfield  {journal}
  {\bibinfo  {journal} {Phys. Rev. Lett.}\ }\textbf {\bibinfo {volume} {122}},\
  \bibinfo {pages} {145901} (\bibinfo {year} {2019})}\BibitemShut {NoStop}%
\bibitem [{\citenamefont {Nomura}\ \emph {et~al.}(2023)\citenamefont {Nomura},
  \citenamefont {Zhang}, \citenamefont {Takagi}, \citenamefont {Karube},
  \citenamefont {Kikkawa}, \citenamefont {Taguchi}, \citenamefont {Tokura},
  \citenamefont {Zherlitsyn}, \citenamefont {Kohama},\ and\ \citenamefont
  {Seki}}]{NomuraPRL2023}%
  \BibitemOpen
  \bibfield  {author} {\bibinfo {author} {\bibfnamefont {T.}~\bibnamefont
  {Nomura}}, \bibinfo {author} {\bibfnamefont {X.-X.}\ \bibnamefont {Zhang}},
  \bibinfo {author} {\bibfnamefont {R.}~\bibnamefont {Takagi}}, \bibinfo
  {author} {\bibfnamefont {K.}~\bibnamefont {Karube}}, \bibinfo {author}
  {\bibfnamefont {A.}~\bibnamefont {Kikkawa}}, \bibinfo {author} {\bibfnamefont
  {Y.}~\bibnamefont {Taguchi}}, \bibinfo {author} {\bibfnamefont
  {Y.}~\bibnamefont {Tokura}}, \bibinfo {author} {\bibfnamefont
  {S.}~\bibnamefont {Zherlitsyn}}, \bibinfo {author} {\bibfnamefont
  {Y.}~\bibnamefont {Kohama}},\ and\ \bibinfo {author} {\bibfnamefont
  {S.}~\bibnamefont {Seki}},\ }\bibfield  {title} {\bibinfo {title}
  {Nonreciprocal phonon propagation in a metallic chiral magnet},\ }\href
  {https://doi.org/10.1103/PhysRevLett.130.176301} {\bibfield  {journal}
  {\bibinfo  {journal} {Phys. Rev. Lett.}\ }\textbf {\bibinfo {volume} {130}},\
  \bibinfo {pages} {176301} (\bibinfo {year} {2023})}\BibitemShut {NoStop}%
\bibitem [{\citenamefont {Fukushima}\ \emph {et~al.}(2024)\citenamefont
  {Fukushima}, \citenamefont {Obata}, \citenamefont {Yamane}, \citenamefont
  {Hu}, \citenamefont {Li}, \citenamefont {Yao}, \citenamefont {Wang},
  \citenamefont {Maeno},\ and\ \citenamefont {Yonezawa}}]{FukushimaNC2024}%
  \BibitemOpen
  \bibfield  {author} {\bibinfo {author} {\bibfnamefont {K.}~\bibnamefont
  {Fukushima}}, \bibinfo {author} {\bibfnamefont {K.}~\bibnamefont {Obata}},
  \bibinfo {author} {\bibfnamefont {S.}~\bibnamefont {Yamane}}, \bibinfo
  {author} {\bibfnamefont {Y.}~\bibnamefont {Hu}}, \bibinfo {author}
  {\bibfnamefont {Y.}~\bibnamefont {Li}}, \bibinfo {author} {\bibfnamefont
  {Y.}~\bibnamefont {Yao}}, \bibinfo {author} {\bibfnamefont {Z.}~\bibnamefont
  {Wang}}, \bibinfo {author} {\bibfnamefont {Y.}~\bibnamefont {Maeno}},\ and\
  \bibinfo {author} {\bibfnamefont {S.}~\bibnamefont {Yonezawa}},\ }\bibfield
  {title} {\bibinfo {title} {\rm{ Violation of emergent rotational symmetry in
  the hexagonal Kagome superconductor CsV$_{3}$Sb$_{5}$}},\ }\href
  {https://doi.org/10.1038/s41467-024-47043-8} {\bibfield  {journal} {\bibinfo
  {journal} {Nat. Commun.}\ }\textbf {\bibinfo {volume} {15}},\ \bibinfo
  {pages} {2888} (\bibinfo {year} {2024})}\BibitemShut {NoStop}%
\bibitem [{\citenamefont {Asaba}\ \emph {et~al.}(2024)\citenamefont {Asaba},
  \citenamefont {Onishi}, \citenamefont {Kageyama}, \citenamefont {Kiyosue},
  \citenamefont {Ohtsuka}, \citenamefont {Suetsugu}, \citenamefont {Kohsaka},
  \citenamefont {Gaggl}, \citenamefont {Kasahara}, \citenamefont {Murayama},
  \citenamefont {Hashimoto}, \citenamefont {Tazai}, \citenamefont {Kontani},
  \citenamefont {Ortiz}, \citenamefont {Wilson}, \citenamefont {Li},
  \citenamefont {Wen}, \citenamefont {Shibauchi},\ and\ \citenamefont
  {Matsuda}}]{AsabaNP2024}%
  \BibitemOpen
  \bibfield  {author} {\bibinfo {author} {\bibfnamefont {T.}~\bibnamefont
  {Asaba}}, \bibinfo {author} {\bibfnamefont {A.}~\bibnamefont {Onishi}},
  \bibinfo {author} {\bibfnamefont {Y.}~\bibnamefont {Kageyama}}, \bibinfo
  {author} {\bibfnamefont {T.}~\bibnamefont {Kiyosue}}, \bibinfo {author}
  {\bibfnamefont {K.}~\bibnamefont {Ohtsuka}}, \bibinfo {author} {\bibfnamefont
  {S.}~\bibnamefont {Suetsugu}}, \bibinfo {author} {\bibfnamefont
  {Y.}~\bibnamefont {Kohsaka}}, \bibinfo {author} {\bibfnamefont
  {T.}~\bibnamefont {Gaggl}}, \bibinfo {author} {\bibfnamefont
  {Y.}~\bibnamefont {Kasahara}}, \bibinfo {author} {\bibfnamefont
  {H.}~\bibnamefont {Murayama}}, \bibinfo {author} {\bibfnamefont
  {K.}~\bibnamefont {Hashimoto}}, \bibinfo {author} {\bibfnamefont
  {R.}~\bibnamefont {Tazai}}, \bibinfo {author} {\bibfnamefont
  {H.}~\bibnamefont {Kontani}}, \bibinfo {author} {\bibfnamefont {B.~R.}\
  \bibnamefont {Ortiz}}, \bibinfo {author} {\bibfnamefont {S.~D.}\ \bibnamefont
  {Wilson}}, \bibinfo {author} {\bibfnamefont {Q.}~\bibnamefont {Li}}, \bibinfo
  {author} {\bibfnamefont {H.~H.}\ \bibnamefont {Wen}}, \bibinfo {author}
  {\bibfnamefont {T.}~\bibnamefont {Shibauchi}},\ and\ \bibinfo {author}
  {\bibfnamefont {Y.}~\bibnamefont {Matsuda}},\ }\bibfield  {title} {\bibinfo
  {title} {\rm{Evidence for an odd-parity nematic phase above the
  charge-density-wave transition in a kagome metal}},\ }\href
  {https://doi.org/10.1038/s41567-023-02272-4} {\bibfield  {journal} {\bibinfo
  {journal} {Nat. Phys.}\ }\textbf {\bibinfo {volume} {20}},\ \bibinfo {pages}
  {40} (\bibinfo {year} {2024})}\BibitemShut {NoStop}%
\bibitem [{\citenamefont {Fang}\ \emph {et~al.}(2025)\citenamefont {Fang},
  \citenamefont {Imamura}, \citenamefont {Mizukami}, \citenamefont {Namba},
  \citenamefont {Ishihara}, \citenamefont {Hashimoto},\ and\ \citenamefont
  {Shibauchi}}]{FangPRL2025}%
  \BibitemOpen
  \bibfield  {author} {\bibinfo {author} {\bibfnamefont {S.}~\bibnamefont
  {Fang}}, \bibinfo {author} {\bibfnamefont {K.}~\bibnamefont {Imamura}},
  \bibinfo {author} {\bibfnamefont {Y.}~\bibnamefont {Mizukami}}, \bibinfo
  {author} {\bibfnamefont {R.}~\bibnamefont {Namba}}, \bibinfo {author}
  {\bibfnamefont {K.}~\bibnamefont {Ishihara}}, \bibinfo {author}
  {\bibfnamefont {K.}~\bibnamefont {Hashimoto}},\ and\ \bibinfo {author}
  {\bibfnamefont {T.}~\bibnamefont {Shibauchi}},\ }\bibfield  {title} {\bibinfo
  {title} {\rm{Field-Angle-Resolved Specific Heat in
  ${\mathrm{Na}}_{2}{\mathrm{Co}}_{2}\mathrm{Te}{\mathrm{O}}_{6}$: Evidence
  against Kitaev Quantum Spin Liquid}},\ }\href
  {https://doi.org/10.1103/PhysRevLett.134.106701} {\bibfield  {journal}
  {\bibinfo  {journal} {Phys. Rev. Lett.}\ }\textbf {\bibinfo {volume} {134}},\
  \bibinfo {pages} {106701} (\bibinfo {year} {2025})}\BibitemShut {NoStop}%
\bibitem [{\citenamefont {Imamura}\ \emph {et~al.}(2025)\citenamefont
  {Imamura}, \citenamefont {Suetsugu}, \citenamefont {Mizukami}, \citenamefont
  {Yoshida}, \citenamefont {Hashimoto}, \citenamefont {Ohtsuka}, \citenamefont
  {Kasahara}, \citenamefont {Kurita}, \citenamefont {Tanaka}, \citenamefont
  {Noh}, \citenamefont {Nasu}, \citenamefont {Moon}, \citenamefont {Matsuda},\
  and\ \citenamefont {Shibauchi}}]{ImamuraSA2025}%
  \BibitemOpen
  \bibfield  {author} {\bibinfo {author} {\bibfnamefont {K.}~\bibnamefont
  {Imamura}}, \bibinfo {author} {\bibfnamefont {S.}~\bibnamefont {Suetsugu}},
  \bibinfo {author} {\bibfnamefont {Y.}~\bibnamefont {Mizukami}}, \bibinfo
  {author} {\bibfnamefont {Y.}~\bibnamefont {Yoshida}}, \bibinfo {author}
  {\bibfnamefont {K.}~\bibnamefont {Hashimoto}}, \bibinfo {author}
  {\bibfnamefont {K.}~\bibnamefont {Ohtsuka}}, \bibinfo {author} {\bibfnamefont
  {Y.}~\bibnamefont {Kasahara}}, \bibinfo {author} {\bibfnamefont
  {N.}~\bibnamefont {Kurita}}, \bibinfo {author} {\bibfnamefont
  {H.}~\bibnamefont {Tanaka}}, \bibinfo {author} {\bibfnamefont
  {P.}~\bibnamefont {Noh}}, \bibinfo {author} {\bibfnamefont {J.}~\bibnamefont
  {Nasu}}, \bibinfo {author} {\bibfnamefont {E.-G.}\ \bibnamefont {Moon}},
  \bibinfo {author} {\bibfnamefont {Y.}~\bibnamefont {Matsuda}},\ and\ \bibinfo
  {author} {\bibfnamefont {T.}~\bibnamefont {Shibauchi}},\ }\bibfield  {title}
  {\bibinfo {title} {\rm{Majorana-fermion origin of the planar thermal Hall
  effect in the Kitaev magnet $\alpha$-RuCl$_{3}$}},\ }\href
  {https://doi.org/10.1126/sciadv.adk3539} {\bibfield  {journal} {\bibinfo
  {journal} {Sci. Adv.}\ }\textbf {\bibinfo {volume} {10}},\ \bibinfo {pages}
  {eadk3539} (\bibinfo {year} {2025})}\BibitemShut {NoStop}%
\bibitem [{\citenamefont {Nakatsuji}\ \emph {et~al.}(2015)\citenamefont
  {Nakatsuji}, \citenamefont {Kiyohara},\ and\ \citenamefont
  {Higo}}]{NakatsujiNature2015}%
  \BibitemOpen
  \bibfield  {author} {\bibinfo {author} {\bibfnamefont {S.}~\bibnamefont
  {Nakatsuji}}, \bibinfo {author} {\bibfnamefont {N.}~\bibnamefont
  {Kiyohara}},\ and\ \bibinfo {author} {\bibfnamefont {T.}~\bibnamefont
  {Higo}},\ }\bibfield  {title} {\bibinfo {title} {Large anomalous hall effect
  in a non-collinear antiferromagnet at room temperature},\ }\href
  {https://doi.org/doi.org/10.1038/nature15723} {\bibfield  {journal} {\bibinfo
   {journal} {Nature}\ }\textbf {\bibinfo {volume} {527}},\ \bibinfo {pages}
  {212} (\bibinfo {year} {2015})}\BibitemShut {NoStop}%
\bibitem [{\citenamefont {Kimata}\ \emph {et~al.}(2019)\citenamefont {Kimata},
  \citenamefont {Chen}, \citenamefont {Kondou}, \citenamefont {Sugimoto},
  \citenamefont {Muduli}, \citenamefont {Ikhlas}, \citenamefont {Omori},
  \citenamefont {Tomita}, \citenamefont {MacDonald}, \citenamefont
  {Nakatsuji},\ and\ \citenamefont {Otani}}]{KimataNature2019}%
  \BibitemOpen
  \bibfield  {author} {\bibinfo {author} {\bibfnamefont {M.}~\bibnamefont
  {Kimata}}, \bibinfo {author} {\bibfnamefont {H.}~\bibnamefont {Chen}},
  \bibinfo {author} {\bibfnamefont {K.}~\bibnamefont {Kondou}}, \bibinfo
  {author} {\bibfnamefont {S.}~\bibnamefont {Sugimoto}}, \bibinfo {author}
  {\bibfnamefont {P.~K.}\ \bibnamefont {Muduli}}, \bibinfo {author}
  {\bibfnamefont {M.}~\bibnamefont {Ikhlas}}, \bibinfo {author} {\bibfnamefont
  {Y.}~\bibnamefont {Omori}}, \bibinfo {author} {\bibfnamefont
  {T.}~\bibnamefont {Tomita}}, \bibinfo {author} {\bibfnamefont {A.~H.}\
  \bibnamefont {MacDonald}}, \bibinfo {author} {\bibfnamefont {S.}~\bibnamefont
  {Nakatsuji}},\ and\ \bibinfo {author} {\bibfnamefont {Y.}~\bibnamefont
  {Otani}},\ }\bibfield  {title} {\bibinfo {title} {Magnetic and magnetic
  inverse spin hall effects in a non-collinear antiferromagnet},\ }\href
  {https://doi.org/10.1038/s41586-018-0853-0} {\bibfield  {journal} {\bibinfo
  {journal} {Nature}\ }\textbf {\bibinfo {volume} {565}},\ \bibinfo {pages}
  {627} (\bibinfo {year} {2019})}\BibitemShut {NoStop}%
\bibitem [{\citenamefont {Qin}\ \emph {et~al.}(2023)\citenamefont {Qin},
  \citenamefont {Yan}, \citenamefont {Wang}, \citenamefont {Chen},
  \citenamefont {Meng}, \citenamefont {Dong}, \citenamefont {Zhu},
  \citenamefont {Cai}, \citenamefont {Feng}, \citenamefont {Zhou},
  \citenamefont {Liu}, \citenamefont {Zhang}, \citenamefont {Zeng},
  \citenamefont {Zhang}, \citenamefont {Jiang},\ and\ \citenamefont
  {Liu}}]{QinNature2023}%
  \BibitemOpen
  \bibfield  {author} {\bibinfo {author} {\bibfnamefont {P.}~\bibnamefont
  {Qin}}, \bibinfo {author} {\bibfnamefont {H.}~\bibnamefont {Yan}}, \bibinfo
  {author} {\bibfnamefont {X.}~\bibnamefont {Wang}}, \bibinfo {author}
  {\bibfnamefont {H.}~\bibnamefont {Chen}}, \bibinfo {author} {\bibfnamefont
  {Z.}~\bibnamefont {Meng}}, \bibinfo {author} {\bibfnamefont {J.}~\bibnamefont
  {Dong}}, \bibinfo {author} {\bibfnamefont {M.}~\bibnamefont {Zhu}}, \bibinfo
  {author} {\bibfnamefont {J.}~\bibnamefont {Cai}}, \bibinfo {author}
  {\bibfnamefont {Z.}~\bibnamefont {Feng}}, \bibinfo {author} {\bibfnamefont
  {X.}~\bibnamefont {Zhou}}, \bibinfo {author} {\bibfnamefont {L.}~\bibnamefont
  {Liu}}, \bibinfo {author} {\bibfnamefont {T.}~\bibnamefont {Zhang}}, \bibinfo
  {author} {\bibfnamefont {Z.}~\bibnamefont {Zeng}}, \bibinfo {author}
  {\bibfnamefont {J.}~\bibnamefont {Zhang}}, \bibinfo {author} {\bibfnamefont
  {C.}~\bibnamefont {Jiang}},\ and\ \bibinfo {author} {\bibfnamefont
  {Z.}~\bibnamefont {Liu}},\ }\bibfield  {title} {\bibinfo {title}
  {Room-temperature magnetoresistance in an all-antiferromagnetic tunnel
  junction},\ }\href {https://doi.org/10.1038/s41586-022-05461-y} {\bibfield
  {journal} {\bibinfo  {journal} {Nature}\ }\textbf {\bibinfo {volume} {613}},\
  \bibinfo {pages} {485} (\bibinfo {year} {2023})}\BibitemShut {NoStop}%
\bibitem [{\citenamefont {Feng}\ \emph {et~al.}(2022)\citenamefont {Feng},
  \citenamefont {Zhou}, \citenamefont {{\AA}~mejkal}, \citenamefont {Wu},
  \citenamefont {Zhu}, \citenamefont {Guo}, \citenamefont
  {Gonz{\~A}{!'}lez-Hern{\~A}{!'}ndez}, \citenamefont {Wang}, \citenamefont
  {Yan}, \citenamefont {Qin}, \citenamefont {Zhang}, \citenamefont {Wu},
  \citenamefont {Chen}, \citenamefont {Meng}, \citenamefont {Liu},
  \citenamefont {Xia}, \citenamefont {Sinova}, \citenamefont {Jungwirth},\ and\
  \citenamefont {Liu}}]{FengNatElectron2022}%
  \BibitemOpen
  \bibfield  {author} {\bibinfo {author} {\bibfnamefont {Z.}~\bibnamefont
  {Feng}}, \bibinfo {author} {\bibfnamefont {X.}~\bibnamefont {Zhou}}, \bibinfo
  {author} {\bibfnamefont {L.}~\bibnamefont {{\AA}~mejkal}}, \bibinfo {author}
  {\bibfnamefont {L.}~\bibnamefont {Wu}}, \bibinfo {author} {\bibfnamefont
  {Z.}~\bibnamefont {Zhu}}, \bibinfo {author} {\bibfnamefont {H.}~\bibnamefont
  {Guo}}, \bibinfo {author} {\bibfnamefont {R.}~\bibnamefont
  {Gonz{\~A}{!'}lez-Hern{\~A}{!'}ndez}}, \bibinfo {author} {\bibfnamefont
  {X.}~\bibnamefont {Wang}}, \bibinfo {author} {\bibfnamefont {H.}~\bibnamefont
  {Yan}}, \bibinfo {author} {\bibfnamefont {P.}~\bibnamefont {Qin}}, \bibinfo
  {author} {\bibfnamefont {X.}~\bibnamefont {Zhang}}, \bibinfo {author}
  {\bibfnamefont {H.}~\bibnamefont {Wu}}, \bibinfo {author} {\bibfnamefont
  {H.}~\bibnamefont {Chen}}, \bibinfo {author} {\bibfnamefont {Z.}~\bibnamefont
  {Meng}}, \bibinfo {author} {\bibfnamefont {L.}~\bibnamefont {Liu}}, \bibinfo
  {author} {\bibfnamefont {Z.}~\bibnamefont {Xia}}, \bibinfo {author}
  {\bibfnamefont {J.}~\bibnamefont {Sinova}}, \bibinfo {author} {\bibfnamefont
  {T.}~\bibnamefont {Jungwirth}},\ and\ \bibinfo {author} {\bibfnamefont
  {Z.}~\bibnamefont {Liu}},\ }\bibfield  {title} {\bibinfo {title} {An
  anomalous hall effect in altermagnetic ruthenium dioxide},\ }\href
  {https://doi.org/10.1038/s41928-022-00866-z} {\bibfield  {journal} {\bibinfo
  {journal} {Nat. Electron.}\ }\textbf {\bibinfo {volume} {5}},\ \bibinfo
  {pages} {735} (\bibinfo {year} {2022})}\BibitemShut {NoStop}%
\bibitem [{\citenamefont {Gonzalez~Betancourt}\ \emph
  {et~al.}(2023)\citenamefont {Gonzalez~Betancourt}, \citenamefont
  {Zub\'a\ifmmode~\check{c}\else \v{c}\fi{}}, \citenamefont
  {Gonzalez-Hernandez}, \citenamefont {Geishendorf}, \citenamefont {\ifmmode
  \check{S}\else \v{S}\fi{}ob\'a\ifmmode~\check{n}\else \v{n}\fi{}},
  \citenamefont {Springholz}, \citenamefont {Olejn\'{\i}k}, \citenamefont
  {\ifmmode~\check{S}\else \v{S}\fi{}mejkal}, \citenamefont {Sinova},
  \citenamefont {Jungwirth}, \citenamefont {Goennenwein}, \citenamefont
  {Thomas}, \citenamefont {Reichlov\'a}, \citenamefont {\ifmmode~\check{Z}\else
  \v{Z}\fi{}elezn\'y},\ and\ \citenamefont {Kriegner}}]{GonzalezPRL2023}%
  \BibitemOpen
  \bibfield  {author} {\bibinfo {author} {\bibfnamefont {R.~D.}\ \bibnamefont
  {Gonzalez~Betancourt}}, \bibinfo {author} {\bibfnamefont {J.}~\bibnamefont
  {Zub\'a\ifmmode~\check{c}\else \v{c}\fi{}}}, \bibinfo {author} {\bibfnamefont
  {R.}~\bibnamefont {Gonzalez-Hernandez}}, \bibinfo {author} {\bibfnamefont
  {K.}~\bibnamefont {Geishendorf}}, \bibinfo {author} {\bibfnamefont
  {Z.}~\bibnamefont {\ifmmode \check{S}\else
  \v{S}\fi{}ob\'a\ifmmode~\check{n}\else \v{n}\fi{}}}, \bibinfo {author}
  {\bibfnamefont {G.}~\bibnamefont {Springholz}}, \bibinfo {author}
  {\bibfnamefont {K.}~\bibnamefont {Olejn\'{\i}k}}, \bibinfo {author}
  {\bibfnamefont {L.}~\bibnamefont {\ifmmode~\check{S}\else \v{S}\fi{}mejkal}},
  \bibinfo {author} {\bibfnamefont {J.}~\bibnamefont {Sinova}}, \bibinfo
  {author} {\bibfnamefont {T.}~\bibnamefont {Jungwirth}}, \bibinfo {author}
  {\bibfnamefont {S.~T.~B.}\ \bibnamefont {Goennenwein}}, \bibinfo {author}
  {\bibfnamefont {A.}~\bibnamefont {Thomas}}, \bibinfo {author} {\bibfnamefont
  {H.}~\bibnamefont {Reichlov\'a}}, \bibinfo {author} {\bibfnamefont
  {J.}~\bibnamefont {\ifmmode~\check{Z}\else \v{Z}\fi{}elezn\'y}},\ and\
  \bibinfo {author} {\bibfnamefont {D.}~\bibnamefont {Kriegner}},\ }\bibfield
  {title} {\bibinfo {title} {\rm{Spontaneous Anomalous Hall Effect Arising from
  an Unconventional Compensated Magnetic Phase in a Semiconductor}},\ }\href
  {https://doi.org/10.1103/physrevlett.130.036702} {\bibfield  {journal}
  {\bibinfo  {journal} {Phys. Rev. Lett.}\ }\textbf {\bibinfo {volume} {130}},\
  \bibinfo {pages} {036702} (\bibinfo {year} {2023})}\BibitemShut {NoStop}%
\bibitem [{\citenamefont {Barzola-Quiquia}\ \emph {et~al.}(2019)\citenamefont
  {Barzola-Quiquia}, \citenamefont {Esquinazi}, \citenamefont {Precker},
  \citenamefont {Stiller}, \citenamefont {Zoraghi}, \citenamefont {F\"orster},
  \citenamefont {Herrmannsd\"orfer},\ and\ \citenamefont
  {Coniglio}}]{BarzolaPRM2019}%
  \BibitemOpen
  \bibfield  {author} {\bibinfo {author} {\bibfnamefont {J.}~\bibnamefont
  {Barzola-Quiquia}}, \bibinfo {author} {\bibfnamefont {P.~D.}\ \bibnamefont
  {Esquinazi}}, \bibinfo {author} {\bibfnamefont {C.~E.}\ \bibnamefont
  {Precker}}, \bibinfo {author} {\bibfnamefont {M.}~\bibnamefont {Stiller}},
  \bibinfo {author} {\bibfnamefont {M.}~\bibnamefont {Zoraghi}}, \bibinfo
  {author} {\bibfnamefont {T.}~\bibnamefont {F\"orster}}, \bibinfo {author}
  {\bibfnamefont {T.}~\bibnamefont {Herrmannsd\"orfer}},\ and\ \bibinfo
  {author} {\bibfnamefont {W.~A.}\ \bibnamefont {Coniglio}},\ }\bibfield
  {title} {\bibinfo {title} {High-field magnetoresistance of graphite
  revised},\ }\href {https://doi.org/10.1103/PhysRevMaterials.3.054603}
  {\bibfield  {journal} {\bibinfo  {journal} {Phys. Rev. Mater.}\ }\textbf
  {\bibinfo {volume} {3}},\ \bibinfo {pages} {054603} (\bibinfo {year}
  {2019})}\BibitemShut {NoStop}%
\bibitem [{\citenamefont {Ikeda}\ \emph {et~al.}(2024)\citenamefont {Ikeda},
  \citenamefont {Noda}, \citenamefont {Shimbori}, \citenamefont {Seki},
  \citenamefont {Bhoi}, \citenamefont {Ishita}, \citenamefont {Nakamura},
  \citenamefont {Matsubayashi},\ and\ \citenamefont {Akiba}}]{IkedaJAP2024}%
  \BibitemOpen
  \bibfield  {author} {\bibinfo {author} {\bibfnamefont {A.}~\bibnamefont
  {Ikeda}}, \bibinfo {author} {\bibfnamefont {K.}~\bibnamefont {Noda}},
  \bibinfo {author} {\bibfnamefont {K.}~\bibnamefont {Shimbori}}, \bibinfo
  {author} {\bibfnamefont {K.}~\bibnamefont {Seki}}, \bibinfo {author}
  {\bibfnamefont {D.}~\bibnamefont {Bhoi}}, \bibinfo {author} {\bibfnamefont
  {A.}~\bibnamefont {Ishita}}, \bibinfo {author} {\bibfnamefont
  {J.}~\bibnamefont {Nakamura}}, \bibinfo {author} {\bibfnamefont
  {K.}~\bibnamefont {Matsubayashi}},\ and\ \bibinfo {author} {\bibfnamefont
  {K.}~\bibnamefont {Akiba}},\ }\bibfield  {title} {\bibinfo {title} {\rm{A
  concise 40 T pulse magnet for condensed matter experiments}},\ }\href
  {https://doi.org/10.1063/5.0231640} {\bibfield  {journal} {\bibinfo
  {journal} {J. Appl. Phys.}\ }\textbf {\bibinfo {volume} {136}},\ \bibinfo
  {pages} {175902} (\bibinfo {year} {2024})}\BibitemShut {NoStop}%
\bibitem [{\citenamefont {Akiba}\ \emph {et~al.}(2015)\citenamefont {Akiba},
  \citenamefont {Miyake}, \citenamefont {Yaguchi}, \citenamefont {Matsuo},
  \citenamefont {Kindo},\ and\ \citenamefont {Tokunaga}}]{AkibaJPSJ2015}%
  \BibitemOpen
  \bibfield  {author} {\bibinfo {author} {\bibfnamefont {K.}~\bibnamefont
  {Akiba}}, \bibinfo {author} {\bibfnamefont {A.}~\bibnamefont {Miyake}},
  \bibinfo {author} {\bibfnamefont {H.}~\bibnamefont {Yaguchi}}, \bibinfo
  {author} {\bibfnamefont {A.}~\bibnamefont {Matsuo}}, \bibinfo {author}
  {\bibfnamefont {K.}~\bibnamefont {Kindo}},\ and\ \bibinfo {author}
  {\bibfnamefont {M.}~\bibnamefont {Tokunaga}},\ }\bibfield  {title} {\bibinfo
  {title} {Possible excitonic phase of graphite in the quantum limit state},\
  }\href {https://doi.org/10.7566/jpsj.84.054709} {\bibfield  {journal}
  {\bibinfo  {journal} {J. Phys. Soc. Jpn.}\ }\textbf {\bibinfo {volume}
  {84}},\ \bibinfo {pages} {054709} (\bibinfo {year} {2015})}\BibitemShut
  {NoStop}%
\bibitem [{\citenamefont {Akiba}\ \emph {et~al.}(2017)\citenamefont {Akiba},
  \citenamefont {Miyake}, \citenamefont {Akahama}, \citenamefont
  {Matsubayashi}, \citenamefont {Uwatoko},\ and\ \citenamefont
  {Tokunaga}}]{AkibaPRB2017}%
  \BibitemOpen
  \bibfield  {author} {\bibinfo {author} {\bibfnamefont {K.}~\bibnamefont
  {Akiba}}, \bibinfo {author} {\bibfnamefont {A.}~\bibnamefont {Miyake}},
  \bibinfo {author} {\bibfnamefont {Y.}~\bibnamefont {Akahama}}, \bibinfo
  {author} {\bibfnamefont {K.}~\bibnamefont {Matsubayashi}}, \bibinfo {author}
  {\bibfnamefont {Y.}~\bibnamefont {Uwatoko}},\ and\ \bibinfo {author}
  {\bibfnamefont {M.}~\bibnamefont {Tokunaga}},\ }\bibfield  {title} {\bibinfo
  {title} {Two-carrier analyses of the transport properties of black phosphorus
  under pressure},\ }\href {https://doi.org/10.1103/PhysRevB.95.115126}
  {\bibfield  {journal} {\bibinfo  {journal} {Phys. Rev. B}\ }\textbf {\bibinfo
  {volume} {95}},\ \bibinfo {pages} {115126} (\bibinfo {year}
  {2017})}\BibitemShut {NoStop}%
\bibitem [{\citenamefont {Ikeda}\ \emph {et~al.}(2017)\citenamefont {Ikeda},
  \citenamefont {Nomura}, \citenamefont {Matsuda}, \citenamefont {Tani},
  \citenamefont {Kobayashi}, \citenamefont {Watanabe},\ and\ \citenamefont
  {Sato}}]{IkedaRSI2017}%
  \BibitemOpen
  \bibfield  {author} {\bibinfo {author} {\bibfnamefont {A.}~\bibnamefont
  {Ikeda}}, \bibinfo {author} {\bibfnamefont {T.}~\bibnamefont {Nomura}},
  \bibinfo {author} {\bibfnamefont {Y.~H.}\ \bibnamefont {Matsuda}}, \bibinfo
  {author} {\bibfnamefont {S.}~\bibnamefont {Tani}}, \bibinfo {author}
  {\bibfnamefont {Y.}~\bibnamefont {Kobayashi}}, \bibinfo {author}
  {\bibfnamefont {H.}~\bibnamefont {Watanabe}},\ and\ \bibinfo {author}
  {\bibfnamefont {K.}~\bibnamefont {Sato}},\ }\bibfield  {title} {\bibinfo
  {title} {\rm{High-speed 100 MHz strain monitor using fiber Bragg grating and
  optical filter for magnetostriction measurements under ultrahigh magnetic
  fields}},\ }\href {https://doi.org/10.1063/1.4999452} {\bibfield  {journal}
  {\bibinfo  {journal} {Rev. Sci. Instrum.}\ }\textbf {\bibinfo {volume}
  {88}},\ \bibinfo {pages} {083906} (\bibinfo {year} {2017})}\BibitemShut
  {NoStop}%
\bibitem [{\citenamefont {Ikeda}\ \emph {et~al.}(2018)\citenamefont {Ikeda},
  \citenamefont {Matsuda},\ and\ \citenamefont {Tsuda}}]{IkedaRSI2018}%
  \BibitemOpen
  \bibfield  {author} {\bibinfo {author} {\bibfnamefont {A.}~\bibnamefont
  {Ikeda}}, \bibinfo {author} {\bibfnamefont {Y.~H.}\ \bibnamefont {Matsuda}},\
  and\ \bibinfo {author} {\bibfnamefont {H.}~\bibnamefont {Tsuda}},\ }\bibfield
   {title} {\bibinfo {title} {\rm{Note: Optical filter method for
  high-resolution magnetostriction measurement using fiber Bragg grating under
  millisecond-pulsed high magnetic fields at cryogenic temperatures}},\ }\href
  {https://doi.org/10.1063/1.5034035} {\bibfield  {journal} {\bibinfo
  {journal} {Rev. Sci. Instrum.}\ }\textbf {\bibinfo {volume} {89}},\ \bibinfo
  {pages} {096103} (\bibinfo {year} {2018})}\BibitemShut {NoStop}%
\bibitem [{\citenamefont {Ikeda}\ \emph {et~al.}(2019)\citenamefont {Ikeda},
  \citenamefont {Furukawa}, \citenamefont {Janson}, \citenamefont {Matsuda},
  \citenamefont {Takeyama}, \citenamefont {Yajima}, \citenamefont {Hiroi},\
  and\ \citenamefont {Ishikawa}}]{IkedaPRB2019}%
  \BibitemOpen
  \bibfield  {author} {\bibinfo {author} {\bibfnamefont {A.}~\bibnamefont
  {Ikeda}}, \bibinfo {author} {\bibfnamefont {S.}~\bibnamefont {Furukawa}},
  \bibinfo {author} {\bibfnamefont {O.}~\bibnamefont {Janson}}, \bibinfo
  {author} {\bibfnamefont {Y.~H.}\ \bibnamefont {Matsuda}}, \bibinfo {author}
  {\bibfnamefont {S.}~\bibnamefont {Takeyama}}, \bibinfo {author}
  {\bibfnamefont {T.}~\bibnamefont {Yajima}}, \bibinfo {author} {\bibfnamefont
  {Z.}~\bibnamefont {Hiroi}},\ and\ \bibinfo {author} {\bibfnamefont
  {H.}~\bibnamefont {Ishikawa}},\ }\bibfield  {title} {\bibinfo {title}
  {Magnetoelastic couplings in the deformed kagome quantum spin lattice of
  volborthite},\ }\href {https://doi.org/10.1103/PhysRevB.99.140412} {\bibfield
   {journal} {\bibinfo  {journal} {Phys. Rev. B}\ }\textbf {\bibinfo {volume}
  {99}},\ \bibinfo {pages} {140412} (\bibinfo {year} {2019})}\BibitemShut
  {NoStop}%
\end{thebibliography}%
\end{document}